\documentclass{article}

\title{A self-consistent solution in affine space with scalar field}

\author{V.Dorofeev\thanks{Friedmann Laboratory For Theoretical Physics,
Department of Mathematics, SPb EF University, Sadovaya 21, 191023
St.Petersburg, Russia, E-mail: dor@vd8186.spb.edu}}

\begin{document}

\maketitle

\begin{abstract}
Conformal connection of scalar field is shown to produce possible
non-metricity in affine connection spaces. In case of
self-consistent solution the non-metricity is a correction to
background Riemannian structure with respect to gravitational
constant and its magnitude may be essential in the early Universe.
\end{abstract}

\section{Introduction}

Riemannian structure of spacetime can be defined axiomatically. On
the other hand, the idea to derive it from a more general setting is
understandable. In this paper we address the following question:
will spacetime possess a Riemannian structure starting from the
simplest presumptions about external fields? One can assume the
metric in self-consistent solution to be induced by the matter---a
scalar field in the Universe. But its Riemannian structure need not
be initially presumed, and the Lagrangian of the geometrical part is
defined by the generalized curvature. Is the field in question
conformal or not---it turns out to be the crucial point in this
consideration and this is the main object of the paper. So, our
basic assumption is that the matter defines the geometrical (not
necessarily Riemannian) structure of spacetime \cite{Dor}.

\section{Self consistent solution for a scalar field}

Consider the full Lagrangian of the form
\begin{equation}\label{curv1}
L=-\frac1{2\kappa}R+L_{\hbox{\footnotesize ext}},
\end{equation}
where $L_{\hbox{\footnotesize ext}}$ is the Lagrangian of matter
fields, $\kappa$ is the gravitational constant and $R$ is the
generalized curvature:
\begin{equation}\label{curv2}
R=g^{ik}(\Gamma^l_{ik,l}-
\Gamma^l_{il,k}+\Gamma^l_{ik}\Gamma^n_{ln}-
\Gamma^n_{il}\Gamma^l_{kn}),
\end{equation}
where $\Gamma^l_{ik}$ is the generalized connection:
\begin{equation}\label{curv3}
\Gamma^l_{ik}=\frac12g^{lp}(g_{pk,l}+g_{pl,k}-g_{kl,p})+T^l_{ik},
\end{equation}
$g^{ik}$ is the spacetime metric and $T^l_{pk}$ is a part of
generalized connection that can not be expressed via the metric
tensor $g^{ik}$.

\medskip

Self-consistent solutions are the solutions of Euler-Lagrange
equations with respect to variables \{$g^{ik},\Gamma^i_{kl}$, matter
fields\}.

\noindent The reasons for non-metric structure of spacetime to occur worth
more detailed study. We assume that the external fields are
exhausted by a massive scalar field with conformal connection $\xi
R$. The the generalized action reads:
\begin{equation}\label{curv4}
S=\int_\Omega(-\frac1{2\kappa}R+\frac12\partial_i\varphi
\partial^i\varphi-\frac12(m^2-\xi R)\varphi^2)\sqrt{-g}d\Omega.
\end{equation}

\noindent Let us find the variation of each variable in (\ref{curv4}).
Introduce $\Lambda=-\frac1{2\kappa}+\frac12\xi R\varphi^2$ and
consider $\delta\Gamma^i_{kl}$.

$$\delta S_1=-\int\{\Lambda[-
g^{ik}(\delta^n_i\delta^m_k\Gamma^p_{lp}+
\delta^m_l\Gamma^n_{ik}-\delta^n_k\Gamma^m_{il}-
\delta^n_i\Gamma^m_{kl})+ g^{nm}_{,l}-g^{np}_{,p}\delta^m_l-$$
\begin{equation}\label{curv5}
-\frac12g^{nm}g_{st}g^{st}_{,l}+
\frac12g^{np}\delta^m_lg_{st}g^{st}_{,p}]+g^{nm}\Lambda_{,l}
-g^{np}\delta^m_l\Lambda_{,p}\}\delta\Gamma^l_{nm}\sqrt{-
g}d\Omega.
\end{equation}

\noindent The values $\delta g^{ik},\delta\varphi$ and $\delta\Gamma^i_{ik}$
are independent on the equations of motion, therefore $\delta S_1=0$
and the expression in braces must vanish. It turns out that the
affine connection is not accorded with the metric and we thus
introduce the non-metricity tensor $T^i_{kl}$:
$$\Gamma^i_{kl}=\tilde\Gamma^i_{kl}+T^i_{kl},$$
\noindent where
$$\tilde\Gamma^i_{kl}=\frac12g^{ip}(g_{pk,l}+g_{pl,k}+g_{kl,p}).$$

\noindent Then the term with $\tilde\Gamma^i_{kl}$ in (\ref{curv5}) vanishes,
and the remaining terms satisfy the following equations:
\begin{equation}\label{curv6}
\Lambda(-g^{nm}T^p_{lp}-\delta^m_lg^{ik}T^n_{ik}+2g^{in}T^m_{il})+
g^{nm}\Lambda_{,l}-g^{np}\delta^m_p\Lambda_{,p}=0.
\end{equation}

\noindent Let
\begin{equation}\label{curv7}
a^m_{kl}=(\delta^m_l\Lambda_{,k}-\delta^m_k\Lambda_{,l})/\Lambda.
\end{equation}

\noindent Then
$2T^m_{kl}=\delta^m_lT^p_{lp}+\delta^m_lg_{nk}g^{pq}T^n_{pq}+a^m_{
kl}$.

\noindent Multiplying (\ref{curv7}) by $g^{kl}$, then by $\delta^l_m$ and
summing up the results we get
\begin{equation}\label{curv8}
T^i_{kl}=\frac12a^i_{kl}-
\frac16(\delta^i_ka^p_{lp}+\delta^i_la^p_{kp}).
\end{equation}

\noindent So, we get
\begin{equation}\label{curv9}
\Gamma^i_{kl}=\tilde\Gamma^i_{kl}-\delta^i_k\Lambda_{,l}/\Lambda.
\end{equation}

\noindent As a result, we find the Ricci tensor
\begin{equation}\label{curv10}
R_{ik}=\tilde R_{ik}+\tilde\Gamma^m_{im}\Lambda_{,m}/\Lambda
\end{equation}
\noindent and the generalized curvature
\begin{equation}\label{curv11}
R=\tilde R+(g^{ik}\tilde\Gamma^m_{ik}\Lambda_{,m}-
\tilde\Gamma^m_{im}\Lambda^{,i})\Lambda.
\end{equation}

\noindent The remaining equations read:
$$\partial_i\varphi\partial_k\varphi-\frac12g_{ik}
(\partial_p\varphi\partial^p\varphi-m^2\varphi^2)=\Lambda(R_{ik}-
\frac12g_{ik}R),$$
$$\nabla_i\nabla^i\varphi+(m^2-\xi R)\varphi=0,$$
where $\nabla_i$ stands for covariant derivative.

\section{Non-metricity in self-consistent solutions}

Recall that
$$\Lambda=-\frac1{2\kappa}+\frac12\xi\varphi^2=
-\frac1{2\kappa}(1-\kappa\xi\varphi^2).$$

\noindent However,
$$\Lambda_{,l}/\Lambda\approx-\kappa\xi(\varphi^2)_{,l}.$$
since the gravitational constant is very small on time scales
different from Plankean $|\kappa\xi R\varphi^2\ll1|$. Therefore the
torsion has the magnitude of the order of gravitational constant,
thus the perturbation theory is applicable:
$$\Gamma^i_{kl}\approx\tilde\Gamma^i_{kl}+\kappa\xi\delta^i_k(\varphi^2)_{,l}$$

On the other hand, suppose there is a domain where the matter is
highly inhomogeneous and $\varphi_{,l}$ is large, that is, of order
of the speed of the fields (for example, near black holes). Then the
additional term will give a big impact (note that the series
expansion with respect to $\kappa$  still takes place as the
expansion is carried out with respect to the field rather than to
the speed) and
$|\tilde\Gamma^i_{kl}|\approx|\kappa\xi\delta^i_k(\varphi^2)_{,l}
|$. It worth mentioning that the impact of the scalar field to
non-metricity is exclusively due to the conformal connection, and
the self-consistent solution contains no non-metricity when $\xi=0$.

\medskip

If the metric in the Universe is conformally invariant
$g_{ik}=a^2(\tau)\eta_{ik}$ and $\tilde\Gamma^l_{il}=4a_{,i}/a$
($\tau$ is conformal time), then
$$R_{ik}=\frac{4a_{,i}a_{,k}-g_{ik}a_{,l}a^{,l}-
g_{ik}a^{,l}_{,l}a-2aa_{,i,k}}{a^2}
+\frac{a_{,k}\Lambda_{,k}-3a_{,i}\Lambda_{,k}-
g_{ik}\Lambda_{,l}a^{,l}}{\Lambda a}$$
$$R=-6\frac{a^{,l}_{,l}}a-6\frac{a^{,l}\Lambda_{,l}}{a\Lambda}.$$

\noindent In particular, for Friedmann metrics $a=a(t)$:
$$R=\tilde R-6\frac{a^{,0}\Lambda_{,0}}{a\Lambda}=
\tilde R-6\kappa\frac{a'}{a^3}(\varphi^2)'$$ (the derivatives
are taken with respect to conformal time). That is, even a uniform
and isotropic space can possess non-metricity. For large values of
non-metricity this can slow down the expansion of the Universe,
since the main part of the energy will be wasted to non-metricity
rather than to expansion \cite{izv}. It happens because the number
of particles being created completely compensates the speed of
expansions at early times \cite{Grib}. As a result, the
non-metricity can become very large for small $t$ ($mt\ll1$).

The work was carried with the support of the Russian Ministry of
Education (grant E02-3.1-198).


\begin{thebibliography}{99}
\bibitem{Dor}
Dorofeev V., in {\it Procs. of the VII Conference "Problems of
Modern General Relativity".} Yerevan, 1985. (in Russian)

\bibitem{izv}
Riess Adam G. et al. \emph{Type Ia Supernova Discoveries at z>1 From
the Hubble Space Telescope: Evidence for Past Deceleration and
Constraints on Dark Energy Evolution}, Astrophys.J. 607 (2004)
665--687

\bibitem{Grib}
A.A. Grib, S.G. Mamaev, and V. M. Mostepanenko, \emph{Vacuum Quantum
Effects in Strong Fields} (Energoatonizdat, Moscow, 1988)
\end{thebibliography}
\end{document}